\documentstyle[preprint,aps,epsfig]{revtex}
\def\bea{\begin{eqnarray}}
\def\beann{\begin{eqnarray*}}
\def\beq{\begin{equation}}
\def\eea{\end{eqnarray}}
\def\eeann{\end{eqnarray*}}
\def\eeq{\end{equation}}

\newcommand{\bk}{\bbox{k}}

\begin{document}
\draft
\title{Generalized polarizabilities and electroexcitation of the nucleon}
\author{D. Drechsel$^1$, S.S. Kamalov$^1$\thanks{\noindent
Permanent Address: Laboratory of Theoretical Physics, JINR Dubna, Head Post
Office Box 79, SU-101000 Moscow, Russia},  G. Krein$^2$, B. Pasquini$^1$ and
L. Tiator$^1$ 
}
\address{
1. Institut f\"{u}r Kernphysik, Universit\"{a}t Mainz, 55099 
Mainz, Germany \\
2. Instituto de F\'{\i}sica Te\'{o}rica, Universidade Estadual Paulista \\ 
Rua Pamplona 145, 01405-900 S\~{a}o Paulo, SP - Brazil 
}

\maketitle
\begin{abstract}
  Generalized nucleon polarizabilities for virtual photons can be
  defined in terms of electroproduction cross sections as function of
  the 4-momentum transfer $Q^2$. In particular, the sum of the
  generalized electric and magnetic polarizabilities $\Sigma=\alpha
  +\beta$ and the spin polarizability $\gamma$ can be expressed by
  virtual photon absorption cross sections integrated over the
  excitation energy. These quantities have been calculated within the
  framework of the recently developed unitary isobar model for pion
  photo- and electroproduction on the proton, which describes the
  available experimental data up to an excitation energy of about 1
  GeV. Our results have been compared to the predictions of chiral
  perturbation theory. 
\end{abstract}
\vspace{0.50cm}
\noindent{PACS NUMBERS: 13.60.Fz,11.55.Fv,14.20.Dh,13.60.Le,13.88.+e }

\vspace{0.3cm}
\newpage
\section{Introduction}
\label{sec1}

Our knowledge about the nucleon's ground state and its
electroexcitation spectrum is largely due to experiments with
electromagnetic probes.  The response of the internal degrees of
freedom of the nucleon to an external electromagnetic field can be
described in terms of structure-dependent polarizabilities. For real
photons, these ground-state properties and polarizabilities can be
related to integrals over photoabsorption cross sections by sum rules,
which are based on general principles of physics such as relativity,
causality and unitarity. One of the most prominent examples is the
Gerasimov-Drell-Hearn (GDH) sum rule~\cite{GDH}, which provides an
astounding relationship between the anomalous magnetic moment $\kappa$
of the nucleon and the difference between photoabsorption cross
sections for parallel and antiparallel alignments of the photon and
nucleon helicities, $\sigma_{3/2}(\omega)~-~\sigma_{1/2}(\omega)$.  Closely
related to this sum rule is the forward spin polarizability~$\gamma$, which
involves an integral over the same combination of cross sections, but
weighted by an additional factor of $\omega^{-2}$~\cite{GGT}. Another
example is Baldin's sum rule~\cite{Baldin}, which expresses the sum of the
electric and magnetic polarizabilities, $\Sigma~=~\alpha~+~\beta$, by
an integral over the total photoabsorption cross section,
$\sigma_{tot}(\omega)~=~[\sigma_{3/2}(\omega)~+~\sigma_{1/2}(\omega)]/2$.

The use of virtual photons from electron scattering processes provides
us with even more detailed information on the structure of the
nucleon. In particular, as we increase the four-momentum of the
virtual photon $Q^2$ from the real-photon point ($Q^2=0$) to large
values of $Q^2$, we can investigate the transition from the
nonperturbative to the perturbative regime of quantum chromodynamics
(QCD). Therefore, the generalizations of real-photon sum rules to
virtual photons provide an interesting possibility to study this
transition and the varying role of the relevant degrees of freedom.

The general sum rules may be defined in terms of electroproduction cross
sections as 

\bea
\Sigma(Q^2) &=&
\frac{1}{2\pi^2}\int^{\infty}_{\omega_{th}}\frac{\sigma_T(\omega,Q^2)}
{\omega^2} d\omega,\label{SigQ2} \\
\gamma(Q^2) &=& \frac{1}{4\pi^2}\int^{\infty}_{\omega_{th}}
\frac{\sigma_{1/2}(\omega,Q^2)-\sigma_{3/2}(\omega,Q^2)}{\omega^3} d\omega\ ,
\label{gamQ2}
\eea
where $\omega_{th}=m_{\pi}+(m^2_{\pi}+Q^2)/2M$ is the threshold energy
in the $lab$ frame, $\sigma_T(\omega,Q^2)$ the transverse cross
section of unpolarized electroexcitation, and
$\sigma_{3/2}(\omega,Q^2)$ and $\sigma_{1/2}(\omega,Q^2)$ are the
cross sections for the scattering of polarized electrons on polarized
nucleons with parallel and antiparallel alignments of the electron and
nucleon helicities. In a recent contribution, Edelmann, Kaiser, Piller
and Weise (EKPW)~\cite{EKPW} have evaluated such generalized sum rules
in the framework of the one-loop approximation to (relativistic)
chiral perturbation theory (ChPT)~\cite{relChPT}, supplemented by tree
graphs for the excitation of the $\Delta(1232)$ resonance in the
relativistic Rarita-Schwinger formalism.

Unfortunately, a direct measurement of the generalized
polarizabilities for $Q^2\neq0$ is extremely difficult if not
impossible, because it requires the extraction of the two-photon
exchange contribution to elastic electron-proton scattering~\cite{DR},
the so-called ``dispersion correction''. On the other hand, the
polarizabilities of the nucleon at $Q^2=0$ can be directly measured by
Compton scattering. They appear as the leading structure-dependent
effect in an expansion of the six independent Compton amplitudes for
small photon energies $\omega$. However, also real Compton scattering
is not at all an easy experiment, and therefore the sum rule value of
$\alpha+\beta=\Sigma (Q^2=0)$ provides a useful constraint to
determine the electric ($\alpha)$ and magnetic $(\beta)$
polarizabilities. At present we have only scarce experimental
information on the 4 spin polarizabilities $\gamma_1$ to $\gamma_4$,
except for the sum rule prediction of the ``forward spin
polarizability'' $\gamma_1-\gamma_2-2\gamma_4=\gamma (Q^2=0)$, while a
complete determination of these observables will require the
scattering of polarized photons off polarized protons.

Recently, there have been extensive experimental and theoretical
investigations of virtual Compton scattering (VCS) by means of the
reaction $e+p\rightarrow e'+p'+\gamma$. This process involves the
absorption of a virtual photon and the emission of a real one. In the
limit of long-wave real photons, this process can also be parametrized
in terms of 6 generalized polarizabilities $P_i(Q^2)$~\cite{Gui,Sch}.
However, the reader should note that VCS is characterized by a
transition from a virtual $(Q^2>0)$ to a real $(Q^2=0)$ photon, while in
the context of our present investigation a generalized polarizability
refers to the scattering of a virtual photon with the same 4-momentum
transfer $Q^2$ in the initial and final states.

The generalized polarizabilities of Eqs.~(\ref{SigQ2})
and~(\ref{gamQ2}) are, of course, constrained by real Compton
scattering at $Q^2=0$. Their evolution with increasing values of $Q^2$
is of considerable interest for our understanding of the underlying
dynamics, because these polarizabilities provide severe constraints to
models of the nucleon at low and moderate momentum transfer. With such
a perspective, we shall apply the recently developed Unitary Isobar
Model (UIM) for electroproduction~\cite{DHKT} to investigate these sum
rules in the resonance region. The UIM is based on an effective
phenomenological Lagrangian for Born terms and vector meson exchange
in the t channel (``background'') and the dominant resonances up to the third 
resonance region. For each partial wave the multipoles satisfy the constraints 
of gauge invariance and unitarity, and in the real photon case the results 
agree well with the predictions of dispersion theory~\cite{HDT}. 
The model is able to describe the correct energy dependence of the multipoles 
for photon energies up to $\omega\simeq 1$~GeV, and it provides a good 
description of all experimentally measured differential cross sections and
polarization observables.

The UIM was used to calculate the spin structure functions $g_1$
and $g_2$ in the resonance region for small and intermediate momentum
transfer~\cite{UIMstruct}. The results agree well with the asymmetries
and the spin structure functions recently measured at
SLAC~\cite{E143}.  Moreover, the first moments of the calculated spin
structure functions $g_1$ and $g_2$ fulfill the Gerasimov-Drell-Hearn
and Burkhardt-Cottingham~\cite{BC} sum rules within 5 to 10\%. One of
the striking features of the generalized GDH integral is its rapid
fluctuation with $Q^2$ and in particular a change of sign at $Q^2
\simeq 0.5$~(GeV)$^2$, which imposes severe constraints on any model
for the nucleon structure.  This zero-crossing separates the region
dominated by resonance-driven coherent processes from a region of
essentially incoherent scattering off the nucleon's constituents. A
similar zero-crossing is predicted by ChPT for the generalized spin
polarizability, $\gamma(Q^2)$~\cite{EKPW}, while we shall show that
the UIM excludes such a cross-over for $Q^2\leq $(1 GeV)$^2$.

In the next section we review the basic elements of the UIM, set the
notation and definitions for cross sections and generalized
polarizabilities.  Our results are compared with ChPT in
Section~\ref{sec3}, and our conclusions are presented in
Section~\ref{sec4}.  Finally, some details of the formalism are given
in the Appendix.

\section{Cross Sections and the Generalized Polarizabilities}
\label{sec2}

In this section we set the notation and summarize the main ingredients
of the UIM~\cite{DHKT}. Let $E$ and $E'$ denote the energies of the
electron in the initial and final states in the {\em lab} frame and
$Q^2=-k^2 >0$ the four-momentum of the virtual photon. The
polarization vector of the target nucleon has the components $P_z$ (in
the direction of the three-momentum of the virtual photon $\bk$) and
$P_x$ (perpendicular to $\bk$, in the scattering plane of the electron
and in the half-plane of the outgoing electron).  The differential
cross section for exclusive electroproduction is then expressed in
terms of the 4 virtual photoabsorption cross sections $\sigma_T$,
$\sigma_L$, $\sigma'_{LT}$ and $\sigma'_{TT}$ by~\cite{Dre94}
\beq
\frac{d\sigma}{d\Omega\ dE'} = \Gamma \sigma(\omega,Q^2)\,,
\label{eq1}
\eeq
where
\beq
\sigma = \sigma_T + \epsilon\sigma_L + hP_x\sqrt{2\epsilon(1-\epsilon)}\ 
         \sigma'_{LT} + hP_z\sqrt{1-\epsilon^2}\sigma'_{TT}\, ,
\label{eq2}
\eeq
with
\beq
\Gamma = \frac{\alpha}{2\pi^2}\ \frac{E'}{E}\ \frac{K}{Q^2}\ 
         \frac{1}{1-\epsilon} ,
\label{eq3}
\eeq 
the flux of the virtual photon field, $\epsilon$ the transverse photon
polarization, and $\omega = E-E'$ the $lab$ energy of the virtual
photon. As in Ref.~\cite{DHKT}, the flux is defined by the ``photon
equivalent energy'' $K=k_{\gamma}=(W^2-M^2)/2M$, where
$W$ is the total c.m. energy and $M$ the mass of the target nucleon.
We caution the reader that this definition due to Hand~\cite{hand} is
not unique. In particular many authors use the definition
$\tilde{K}(Q^2)=\sqrt{\omega^2+Q^2}$, which has been originally proposed by
Gilman~\cite{Gilman}. While both definitions agree at the real photon point, 
where they describe the $lab$ momentum of the real photon, they
differ in the case of electron scattering. Since the differential
cross section should be independent of the choice of $K$ or
$\tilde{K}$, a change of definition leads to an additional
$Q^2$-dependent factor $\tilde{K}/K=\sqrt{1+Q^2/\omega^2}/(1-Q^2/2M\omega)$ 
for the virtual photon absorption cross sections of Eq.~(\ref{eq1}) and, as 
a consequence, to different definitions of the generalized polarizabilities
of Eqs.~(\ref{SigQ2}) and~(\ref{gamQ2}).

The cross sections $\sigma_{1/2}$ and
$\sigma_{3/2}$ of Eqs.~(\ref{SigQ2}) and (\ref{gamQ2}) are related to the 
virtual photoabsorption cross sections $\sigma_T$ and $\sigma'_{TT}$ by
\bea
\sigma_T     &=& \frac{1}{2}(\sigma_{3/2} + \sigma_{1/2}) , \label{sum} \\
\sigma'_{TT} &=& \frac{1}{2} (\sigma_{3/2}-\sigma_{1/2}) . \label{diff}
\eea
These cross sections can be also expressed in terms of the
standard quark structure functions $F_1$, $g_1$ and $g_2$
depending on the Bjorken scaling variable $x=Q^2/2M\omega$ and
$Q^2$
\bea \sigma_T = \frac{4\pi^2\alpha}{MK}\,F_1\,,\qquad
\sigma'_{TT}= -
\frac{4\pi^2\alpha}{MK}\,(g_1-\frac{Q^2}{\omega^2}\,g_2)\,.
\label{quark} \eea
Consequently, the generalized polarizabilities $\Sigma$ and $\gamma$
defined by Eqs. (1) and (2) can also be written in the form
\bea \Sigma(Q^2)=\frac{8\alpha
M}{Q^4}\int^{x_0}_{0}\frac{x}{1-x}\, F_1(x,Q^2)\,dx\,, \eea
\bea \gamma(Q^2)=\frac{16\alpha
M^2}{Q^6}\int^{x_0}_{0}\frac{x^2}{1-x}\,
[g_1(x,Q^2)-\frac{Q^2}{\omega^2}\,g_2(x,Q^2)]\,dx\,, \eea
where $x_0=Q^2/(2M m_{\pi}+m_{\pi}^2 + Q^2)$ refers to the
inelastic threshold of one-pion production. Note that in the
scaling  regime the structure functions should depend on $x$ only.

In our previous work~\cite{UIMstruct} we generalized the GDH and BC
sum rules as the first moments of the $g_1$ and $g_2$ structure
functions, respectively, i.e by the integrals $I_{1(2)}(Q^2)=(2
M/Q^2)\int g_{1(2)}(x,Q^2)\,dx$. We found that in the resonance
region ($W<2$ GeV) and for small $Q^2$, the single pion production
gives the dominant contribution to these integrals. However, for 
increasing values of $Q^2$, the role of the $\eta$ and multipion
production channels become important. In the present paper
we include these channels as well, following the procedure of
Ref.~\cite{UIMstruct}.

The dominant contributions to the $\sigma_{1/2}$ and $\sigma_{3/2}$
cross sections, related to the single pion electroproduction, can be
obtained by numerical integration of the corresponding differential
cross sections, which are expressed in terms of the standard CGLN
amplitudes $F_1,...,F_4$~\cite{DHKT}.  In the UIM these amplitudes
receive contributions from Born terms, including vector meson
exchange, and nucleon resonances with large photon couplings up to the
third resonance region, i.e. the resonances $P_{33}(1232)$,
$P_{11}(1440)$, $D_{13}(1520)$, $S_{11}(1535)$, $F_{15}(1680)$, and
$D_{33}(1700)$. The expressions for the cross sections $\sigma_{1/2}$
and $\sigma_{3/2}$ in terms of the CGLN amplitudes are given in the
Appendix.

\section{Results and Discussions}
\label{sec3}

Table~I presents the separate contributions of the model ingredients
to the polarizabilities $\Sigma$ and $\gamma$ of proton and neutron at
the real photon point.  The contribution of the dressed $\Delta(1232)$
excitation and its interference with the Born plus $\omega$ plus $\rho$
(background) terms are denoted by ``$\Delta$''. The column
``$P_{11},D_{13},\dots$'' gives the contribution of all resonances
above the $\Delta(1232)$ and their interference with the background
and the $\Delta(1232)$, and so forth for the columns labeled $\eta$ and
multipion.  Finally, the sum of all contributions
is contained in the column ``total'' of the table.

Our results for $\alpha+\beta$ agree well with the existing analysis
of the sum rules (see, e.g., Ref.~\cite{Tolya} for a review),
\bea
\alpha^p + \beta^p & \simeq & (14.3 \pm 0.5) \times 10^{-4} \text{fm}^3,
\label{Tolpro} \\
\alpha^n + \beta^n & \simeq & (15.8 \pm 0.5) \times 10^{-4} \text{fm}^3. 
\label{Tolneu}
\eea
In a more recent evaluation of the sum rule, Babusci et
al.~\cite{Babusci} found somewhat reduced values,
$\alpha^p+\beta^p=13.69\pm0.14$ and $\alpha^n+\beta^n=14.40\pm 0.66$
in units of $10^{-4}$ fm$^3$. Part of the deviations might be
attributed to the contribution of the deep inelastic domain, which is
included in the calculation of Ref.~\cite{Babusci} but not in our
result. We also found a discrepancy of about $10\%$ in the numerical
calculation of the dispersion integral in the threshold region
($\omega_{thr}\le\omega\le 0.2$ GeV).  Using the same set of pion
photoproduction multipoles of Ref.~\cite{Babusci}, we obtained
$(\alpha+\beta)^{thr}_p=1.14$ and $(\alpha+\beta)^{thr}_n=1.70$
instead of $(\alpha+\beta)^{thr}_p=1.25$ and
$(\alpha+\beta)^{thr}_n=1.86$ as quoted in Ref.~\cite{Babusci}.

As may be seen from Table~I, the Born terms are by far the major
contributor to $\alpha+\beta$, followed by the $\Delta(1232)$
resonance and multipion production. Our total values for $\Sigma^p$
and $\Sigma^n$ are similar to those obtained within relativistic
ChPT~\cite{EKPW},
\bea
\Sigma^p_{\text{ChPT}}(0) &=& (5.48+8.23)\times 10^{-4} \,\text{fm}^3 = 
13.71\times 10^{-4} \,\text{fm}^3 \ ,\label{apbpChPT} \\
\Sigma^n_{\text{ChPT}}(0) &=& (8.90+8.23)\times 10^{-4} \,\text{fm}^3 = 
17.13\times 10^{-4} \,\text{fm}^3 \ ,
\label{apbnChPT}
\eea 
with the two terms in the central part of this equation giving the
individual contributions of pion-loop and $\Delta$-pole terms.
Regarding these individual contributions, however, our results differ
considerably. In particular in the case of the proton, our background
contribution is 50~\% larger than the pion-loop contribution of
Ref.~\cite{EKPW}, and our $\Delta$ contribution is only 25~\% of the
value of that reference. In fact we also find a large value for the
$\Delta(1232)$ alone, $\Sigma^p_{\Delta}= 9.8\times 10^{-4}$ fm$^3$
(see Fig.~4(c)).  However, the interference with the background
reduces this value to 2.04$\times 10^{-4}$ fm$^3$ (see Table~I).

In the case of the spin polarizability $\gamma$, there also occurs a
big cancellation between the ``background'' (essentially S-wave pion
production near threshold) and the $\Delta(1232)$, while all other
contributions are found to be extremely small, because of the damping
factor $1/\omega^3$ in the integrand of Eq.~(\ref{gamQ2}). In the
neutron channel, this cancellation is almost complete and $\gamma_n$
is practically zero. Due to the $1/\omega^3$ damping factor we expect
the contributions of the deep inelastic region to be small as well.

As in the case of the GDH sum rule, the spin polarizability $\gamma$
is very sensitive to an exact treatment of the $E_{0+}$
photoproduction multipole in the threshold region~\cite{DKH,DK}.
Moreover, the value of $\gamma$ is almost entirely given by the
contributions of the multipoles $E_{0+}$ and $M_{1+}$, which
contribute with opposite signs.  The predicted values of the UIM,
\bea
\gamma^p &\simeq& -0.6 \times 10^{-4} \, \text{fm}^4, \\
\gamma^n &\simeq& 0  \times 10^{-4} \, \text{fm}^4,
\eea
are much smaller (in absolute value) than the ones obtained from the
SAID multipoles~\cite{San,SAID}, and relativistic chiral perturbation
theory~\cite{EKPW,relChPT}. We note, however, that the results of
relativistic ChPT~\cite{relChPT} are not based on a systematic
expansion in $1/M$. 

We also point out that our value for $\gamma^p$ carries a sign
opposite to the prediction of heavy baryon ChPT~\cite{HBChPTgammas}.
This is a rather intriguing result, since this theory does indeed
provide a systematic expansion in $1/M$. Similarly as in the case of
$\alpha+\beta$, for which the theory obtains a much too large value,
the reason for this shortcoming might be due to large loop corrections
in fourth order $(\epsilon^4)$, which have been neglected in the
$\epsilon^3$ approximation of Ref.~\cite{HBChPTgammas}. We further
record that a recent calculation based on the Chiral Soliton
Model~\cite{CSM} predicted $\gamma^p~=~\gamma^n~=~-0.1\times
10^{-4}$~fm~$^4$.

The values for the spin polarizabilities predicted from the UIM are in
good agreement with the results of Refs.~\cite{DKH,Dre99}, obtained on
the basis of the HDT multipoles~\cite{HDT}. These multipoles are
generated by dispersion relations at fixed $t$, and they provide an
excellent description of the photoproduction data for $\omega\leq
450$~MeV~\cite{photodata}.  In particular, these multipoles are also
in agreement with the low energy theorems~\cite{relChPT}.

Next, we present our results for the generalized
polarizabilities. In Fig.~1, we show the evolution of $\Sigma(Q^2)$
for (a) the proton and (b) the neutron. Clearly seen in the figure are
the large individual contributions of the Born terms and of the $\Delta(1232)$.
It should also be noted that the contribution of multipion production
can not be neglected.

Fig.~2 shows our predictions for $\gamma(Q^2)$ for (a) the proton and
(b) the neutron. As in the case of real photons, the main contributions are
from the Born terms and the $\Delta(1232)$, and there occur large
cancellations between these two contributions. The higher resonances as well 
as $\eta$ and multipion production are quite negligible due to the weighting 
factor $\omega^{-3}$ in the integrand.

In Fig.~3 we compare the predictions of the UIM (solid lines) and
relativistic ChPT (dashed lines) for the generalized polarizabilities.
As can be seen, there are significant, qualitative and quantitative,
differences between the UIM and ChPT predictions. The most striking
difference refers to the slope of the $\gamma(Q^2)$ close to the real
photon point. With increasing values of $Q^2$, the UIM prediction for
the background drops much faster than the $\Delta(1232)$ contribution,
and as a result a steep slope develops for $Q^2 < 0.1$~(GeV)$^2$.
Relativistic ChPT on the other hand, predicts a rather flat behavior
in this region.  The pronounced slope in $\gamma(Q^2)$ observed
in the UIM is due to the interference between background and
$\Delta(1232)$ terms, as we shall explain later.

In Fig.~4, we show the individual contributions of the background
(dotted lines), the $\Delta(1232)$ {\em only} (dashed lines), and the
interference between background and $\Delta(1232)$ (dash-dotted
lines). The sum of these three contributions is given by the solid
line.  The big destructive interference between background and
$\Delta(1232)$ contributions to $\Sigma(Q^2)$ and $\gamma(Q^2)$ is now
immediately visible, and the pronounced slope of $\gamma(Q^2)$ near
the origin is seen to result from the interference term.
 
It is interesting that we do not find a zero-crossing for
$\gamma^p(Q^2)$ and $\gamma^n(Q^2)$ in the range of $Q^2 \leq
1.0$~(GeV)$^2$, while Ref.~\cite{EKPW} predicts such a crossing at
$Q^2 \sim 0.4$~(GeV)$^2$.  In our previous work~\cite{UIMstruct}, we
found that the UIM gives a zero-crossing at $Q^2 \sim 0.5$~(GeV)$^2$
for the GDH integral $I_1$ which is similar to $\gamma$ but with a
weighting factor $\omega^{-1}$ in the integrand and an extra term
proportional to $\sigma'_{LT}$ (whose contribution is small for $Q^2
\sim 0.5$~(GeV)$^2$). The origin of this phenomenon is the
cancellation between the (negative) contribution from the $\Delta$
resonance and the (positive) contributions from the higher resonances
and $\eta$ plus multipion channels. The contribution of the $\eta$ plus
multipion channels becomes more and more important with increasing
$Q^2$, with the eventual result of a positive value for the GDH
integral.  However, as we have pointed out before, the $\eta$ plus
multipion channels are more strongly suppressed in the case of
$\gamma$ than for the GDH integral. Therefore, the zero crossing of
$\gamma$ does not appear at low values of $Q^2$. Numerically we find
that $\gamma^p$ changes sign at $Q^2\sim$~1.4 (GeV)$^2$, and
$\gamma^n$ at $Q^2\sim$ 2~(GeV)$^2$.

As a final remark, we mention that the discussed interference between 
background and $\Delta(1232)$ terms originates from the {\em
dynamical} dressing of the $\gamma N \Delta$ vertex, which is
pictorially shown in Fig.~5. The main mechanism to renormalize the
$\gamma N \Delta$ vertex is diagram~5(b), because this diagram has a
strong imaginary part. As is obvious from the optical theorem, this
imaginary part individually leads to a strong interference term. In our
calculation of pion production such contributions appear naturally in the 
expressions for the differential cross sections. 

\section{Conclusion}
\label{sec4}

We evaluated the generalized Baldin sum rule $\Sigma(Q^2)$ and the spin
polarizability $\gamma(Q^2)$ for small and moderate values of $Q^2$
using the Unitary Isobar Model (UIM)~\cite{DHKT}. Both $\Sigma(Q^2)$
and $\gamma(Q^2)$ are dominated by background and $\Delta(1232)$
resonance contributions. In addition $\Sigma(Q^2)$ also receives
sizable contributions from multipion processes.

Our predictions were compared with a recent calculation in the
framework of relativistic chiral perturbation theory (ChPT). While the
total value of $\Sigma$ agrees well with the result of ChPT and other
phenomenological calculations, we differ from ChPT in the case of
$\gamma(0)$. Although one has to recognize that the predictions of
relativistic ChPT are not based on a consistent $1/M$ expansion, the
agreement with the phenomenological result based on the SAID
multipoles SP97K, has been taken as some assurance of the
convergence of the expansion. However, as was remarked
earlier~\cite{DKH,DK}, these SAID multipoles did not describe the
threshold dependence of the $E_{0+}$ photoproduction multipole but
were at variance with the low energy theorems. We repeat that the
correct threshold behavior of the multipole $E_{0+}$ is extremely
important for both $\gamma$ and the closely related GDH sum rule.

We also found significant, qualitative and quantitative, differences
between the UIM and ChPT predictions for the evolution of $\Sigma$ and
$\gamma$ with momentum transfer. The most important qualitative
difference concerns the absence of the interference between background
and $\Delta(1232)$ resonance in Ref.~\cite{EKPW}. While this
interference does not lead to big effects for the net value of
$\Sigma(Q^2)$, it has a dramatic effect for $\gamma(Q^2)$, in
particular for $Q^2 < 0.1$~(GeV)$^2$. The physical origin for the
interference is the dynamical dressing of the $\gamma N \Delta$
vertex~\cite{KamYang}.

We are looking forward to experimental tests of our predictions by
polarized electroproduction cross sections, which will become
available in the near future.  There are indeed quite a few proposals
in several laboratories throughout the world to perform such
experiments, e.g. at Jefferson Lab, ELSA and MAMI. The experimental 
data presently available do not allow a very precise determination of
many ingredients of the UIM. Among the most important ones, is the relative
contribution of background and resonances to the multipoles.
While the pseudovector Born terms are well described, additional background
contributions are model dependent, such as loop effects, pion rescattering
or u-channel resonances. The common feature of such effects is that they
are weakly energy dependent and visible mostly in S waves. As far as the 
existing data are concerned, they are well described by the UIM, which is
constrained by unitarity and gauge invariance. Therefore, the UIM should 
provide a reasonable first estimate for the sum rules. Of course, it is only 
with the availability of new experimental data in the near future that models
like the UIM can be firmly tested.  Such data will certainly enhance our 
knowledge on various aspects of nonperturbative QCD in general and, in 
particular, on the low energy spin structure of the nucleon.

\acknowledgments

This work has been supported in part by Deutsche
Forschungsgemeinschaft, SFB 443 (Germany) and by CNPq (Brazil).

\appendix
\section*{}

In this appendix we give the expressions of the single-pion
electroproduction cross sections $\sigma_{1/2}$, $\sigma_{3/2}$
and $\sigma'_{LT}$ in terms of the standard CGLN amplitudes
$F_1,...,F_6$. The definition of these amplitudes is the same as in
Ref.\cite{DHKT}. Within the UIM, they can be calculated using the 
on-line version of the numerical program MAID accessible on the internet
by {\it http://www.kph.uni-mainz.de/T/maid/}.

The expressions for the cross sections are greatly simplified
by introducing the spin amplitudes ${\cal H}_1,...,{\cal H}_6$~\cite{KDT} 
\beann {\cal H}_1 = \frac{-\sin{\theta_{\pi}}}{\sqrt{2}}\, (F_3 +
F_4 \cos{\theta_{\pi}})\,,\qquad
{\cal H}_2 = \frac{-1}{\sqrt{2}}\,(2 F_1 - 2 F_2
\cos{\theta_{\pi}}) + {\cal H}_3\,, \eeann
\beq {\cal H}_3 = \frac{-1}{\sqrt{2}}\,F_4
\sin^2{\theta_{\pi}}\,,\qquad
{\cal H}_4 = \frac{\sin{\theta_{\pi}}}{\sqrt{2}}\, (2 F_2 +  F_3 +
F_4 \cos{\theta_{\pi}})\,, \eeq
\beann {\cal H}_5 = F_5 + F_6 \cos{\theta_{\pi}} \,,\qquad
{\cal H}_6 = F_6 \sin{\theta_{\pi}}\,, \eeann
where $\theta_{\pi}$ is the polar angle of the outgoing pion. In terms of 
these spin amplitudes, the cross sections $\sigma_{1/2}$, 
$\sigma_{3/2}$ and $\sigma'_{LT}$ are given by
\beq
\sigma_{1/2}=\frac{q}{k_{\gamma}^{cm}}\int d\Omega_{\pi}
(\mid {\cal H}_2 \mid^2 + \mid {\cal H}_4 \mid^2)\,,
\eeq
\beq
\sigma_{3/2}=\frac{q}{k_{\gamma}^{cm}}\int d\Omega_{\pi}
(\mid {\cal H}_1 \mid^2 + \mid {\cal H}_3 \mid^2)\,,
\eeq
\beq \sigma'_{LT}=\frac{q}{k_{\gamma}^{cm}} \frac{Q}{\omega^{cm}}
\int d\Omega_{\pi} \frac{1}{\sqrt{2}}Re({\cal H}_5{\cal H}_2^*
+{\cal H}_6{\cal H}_4^*)\,, \eeq
where $q=\mid \bbox{q} \mid$ and $\omega^{cm}=(W^2-M^2-Q^2)/2W$ are
the pion momentum and the virtual photon energy respectively, in the
c.m. frame. The ``photon equivalent energy'' in the c.m. frame is
defined as $k_{\gamma}^{cm}=(W^2-M^2)/2W$. Note that in comparison
with the standard nomenclature of deep inelastic scattering
(DIS)\cite{E143}, our interference cross sections are
$\sigma'_{LT}=-\sigma_{LT}$(DIS) and
$\sigma'_{TT}=-\sigma_{TT}$(DIS)$=(\sigma_{3/2}-\sigma_{1/2})/2$.

\begin{table}[htbp]
\caption{Contributions to 
  $\Sigma(0)=\alpha+\beta$ (in units of $10^{-4}$~fm$^3$) and
  $\gamma(0)=\gamma$ (in units of $10^{-4}$~fm$^4$) for proton and
  neutron. For details see text.}
\begin{center}
\begin{tabular}{lrrrrrr}
${}$ &Born+$\omega+\rho$ & $\Delta$  & $P_{11},D_{13},...$    
& $\eta$ & multipion & total \\
\hline
$\alpha^p+\beta^p$ & 9.17   & 2.04 & 0.56    & 0.08   & 1.56   & 13.41  \\
$\gamma^p$         & 0.90   &-1.51 &-0.03    & 0.02   & -0.03  & -0.65 \\
\\
$\alpha^n+\beta^n$ &10.86   & 2.04  & 0.45   & 0.08   & 1.56   & 14.99  \\
$\gamma^n$         & 1.54   &-1.51  & 0.06   & 0.02   &-0.03   &  0.08 \\
\end{tabular}
\end{center}  
\end{table}

\begin{figure}
\centerline{
{\epsfxsize=8.5cm\epsfbox{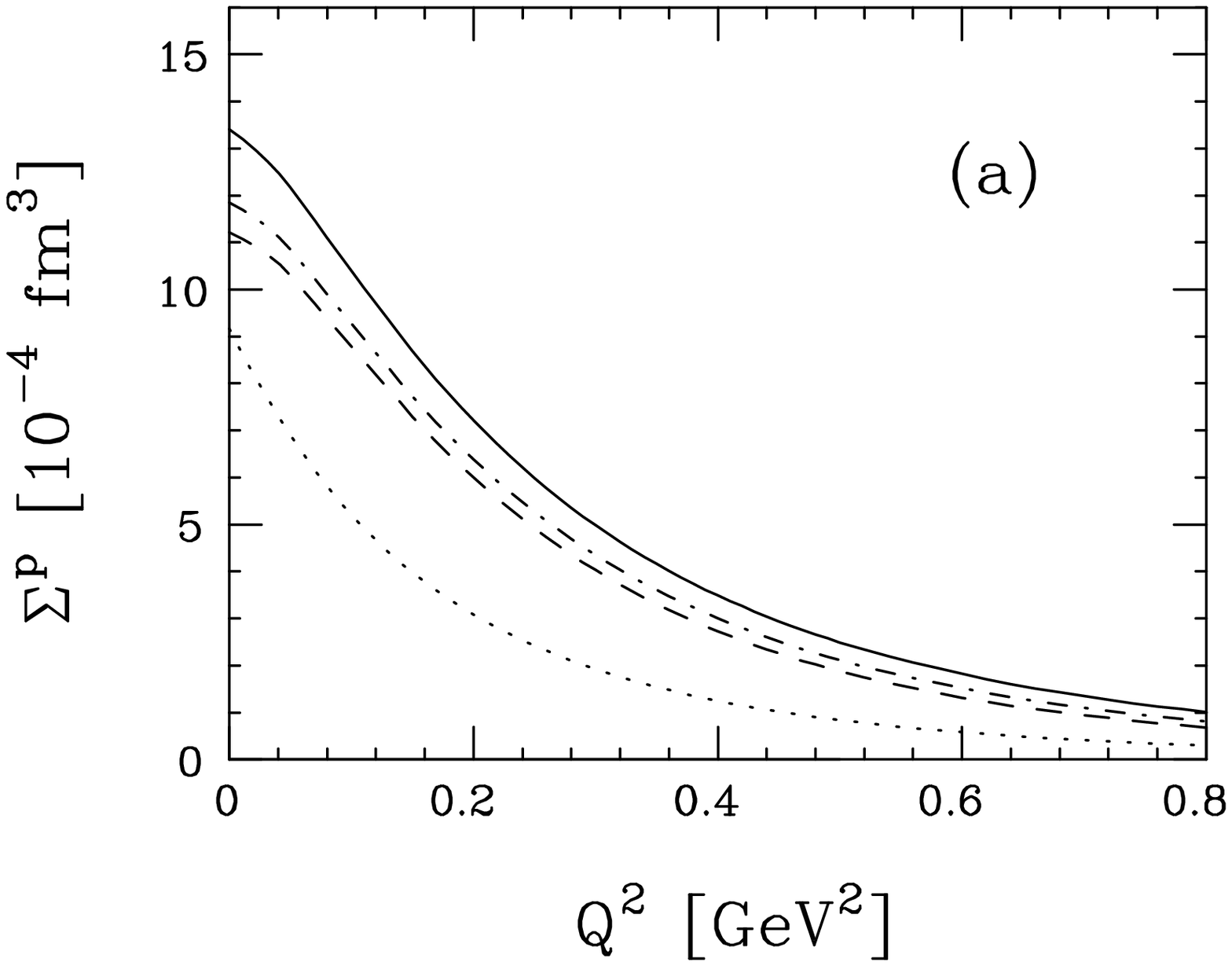}}
{\epsfxsize=8.5cm\epsfbox{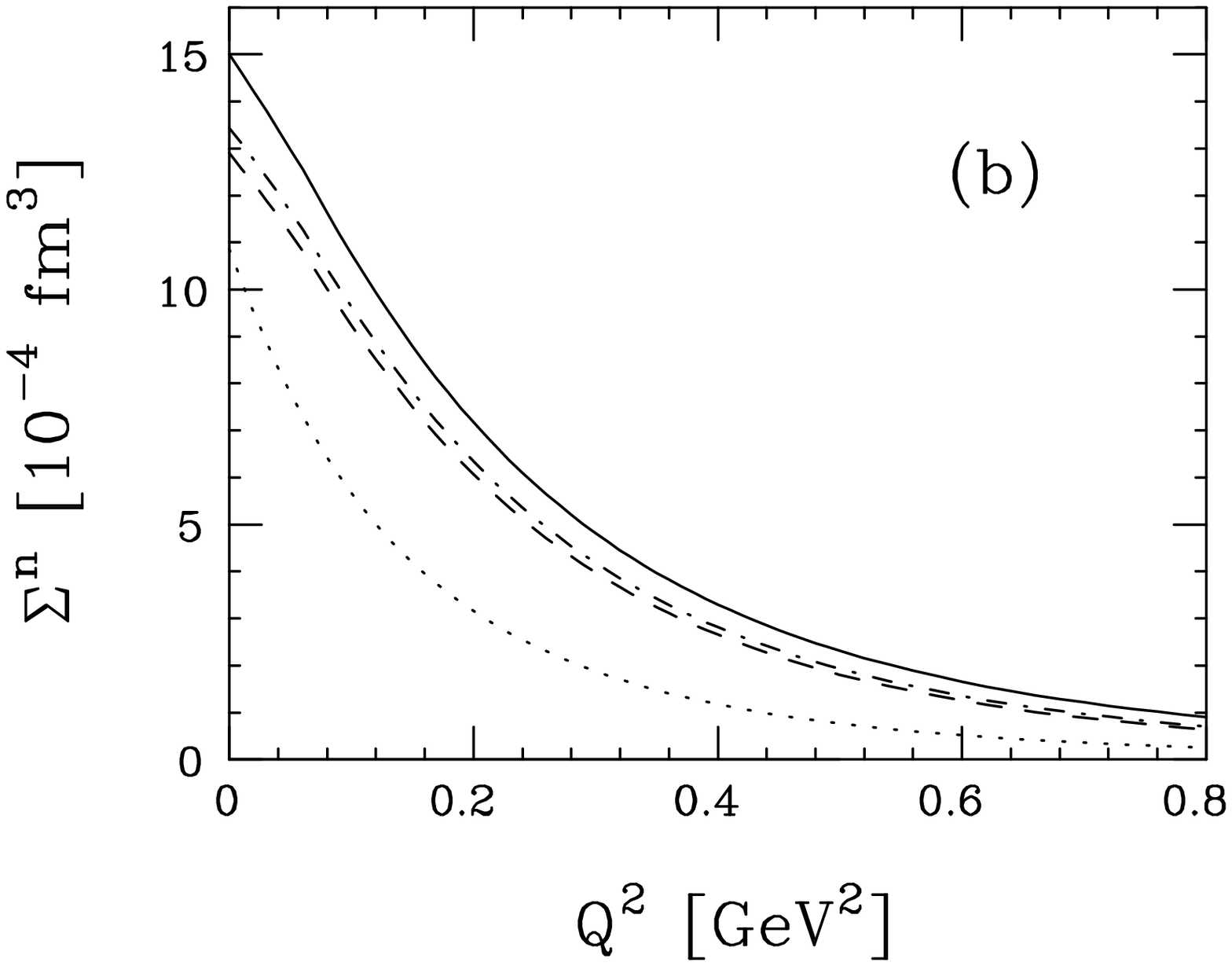}}
}
\vspace{-2.0cm}
\caption{The sum of the electric and magnetic polarizabilities of (a) 
  proton and (b) neutron as function of $Q^2$. The solid line is the
  full result including the $1\pi$, $\eta$ and $n\pi$
  contributions. The dotted line represents the contribution of Born
  terms and vector mesons. The dashed line is obtained by adding the
  $\Delta(1232)$ resonance, and the dash-dotted line by adding all
  resonances and the $\eta$ channel. The difference between the full
  and the dash-dotted lines is therefore due to the production of two
  and more pions.}
\end{figure}

\begin{figure}
\centerline{
{\epsfxsize=8.5cm\epsfbox{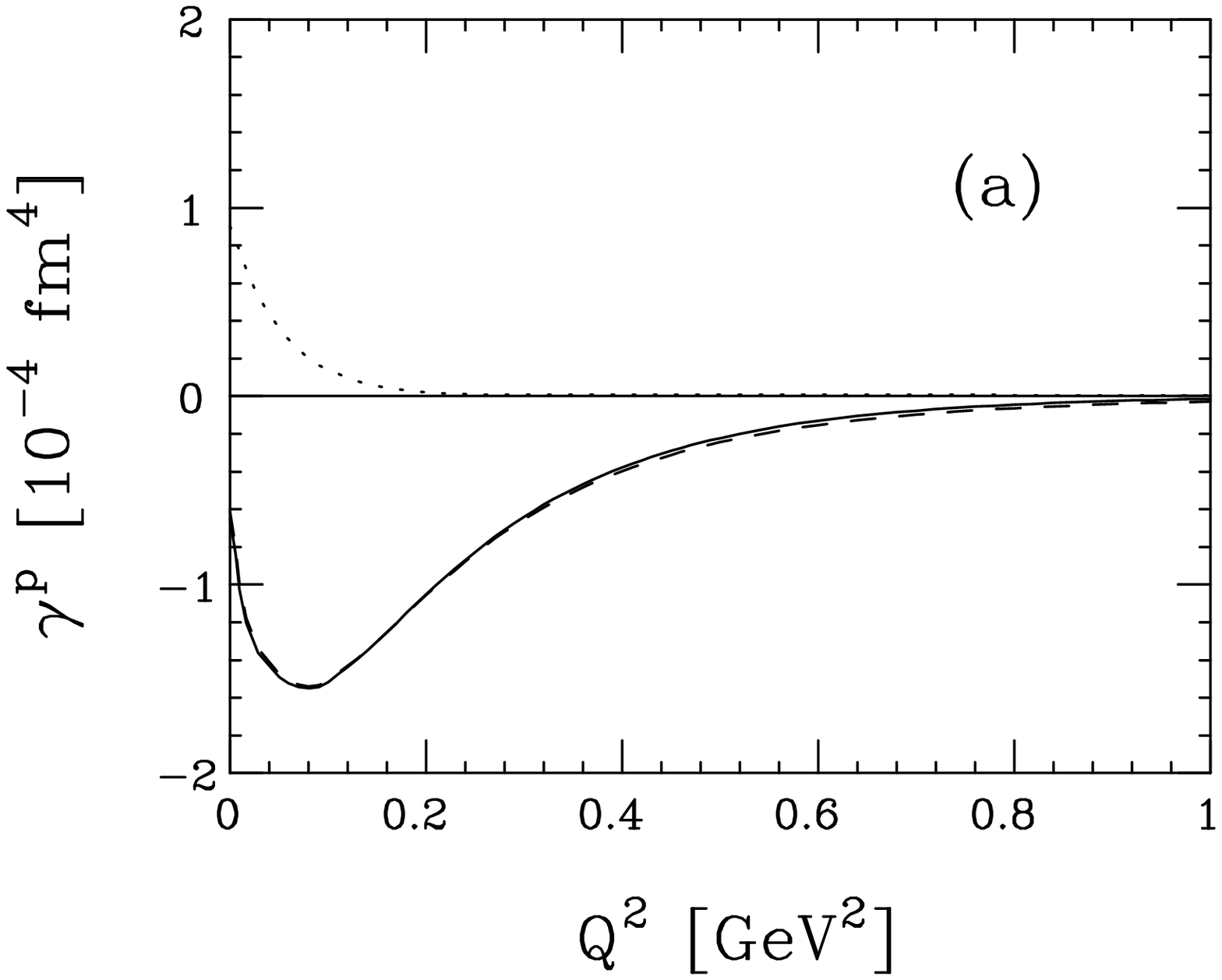}}
{\epsfxsize=8.5cm\epsfbox{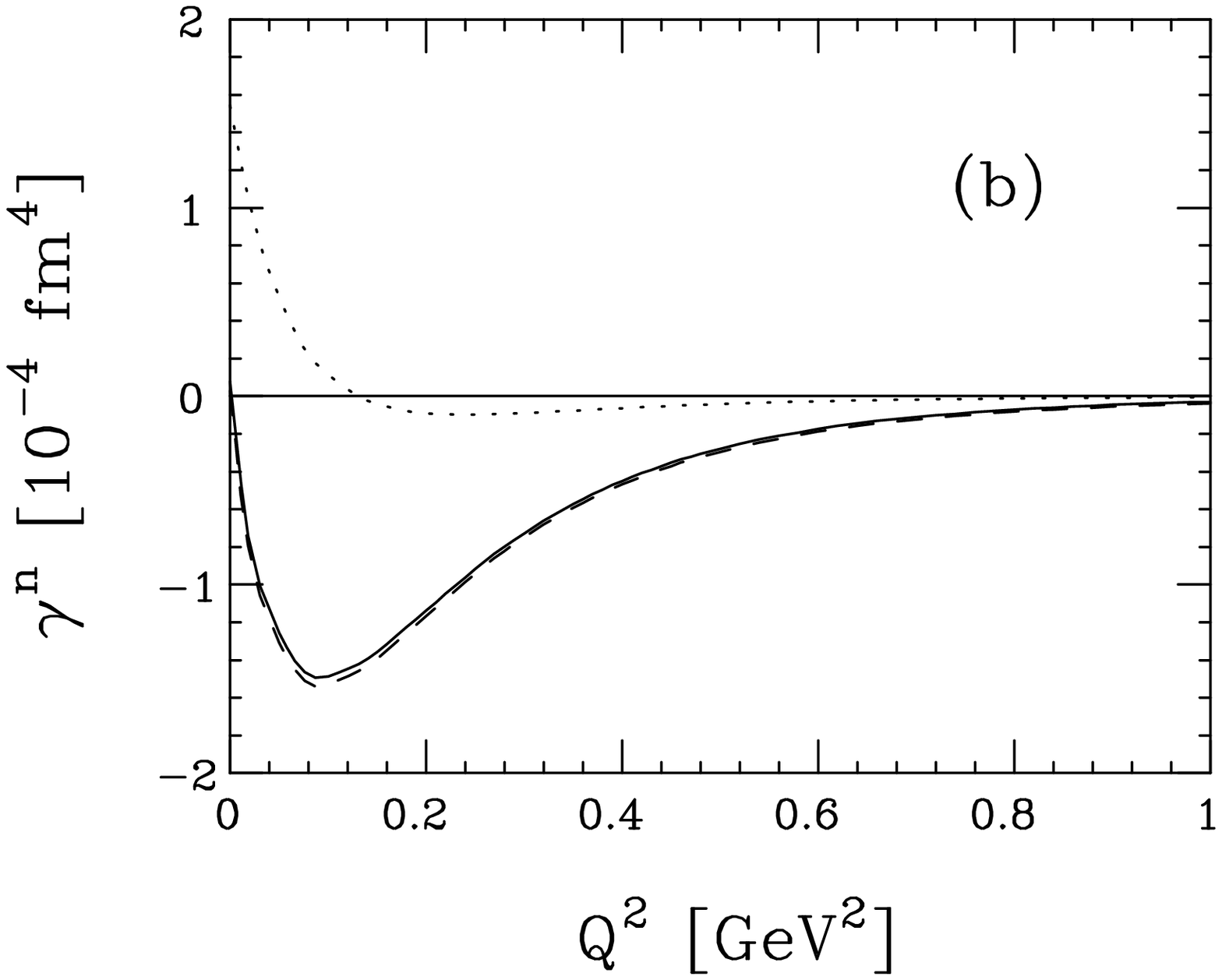}}
}
\vspace{-2.0cm}
\caption{The generalized spin polarizability of (a) proton and (b) neutron
  as function of~$Q^2$. The dotted line represents the contributions
  of Born terms and vector mesons, the dashed line includes both
  the Born terms and the $\Delta(1232)$. The solid line (almost on
  top of the dashed line) also includes the higher resonances as well
  as $\eta$ and multipion production.}
\end{figure}

\newpage
\begin{figure}
\centerline{
{\epsfxsize=8.5cm\epsfbox{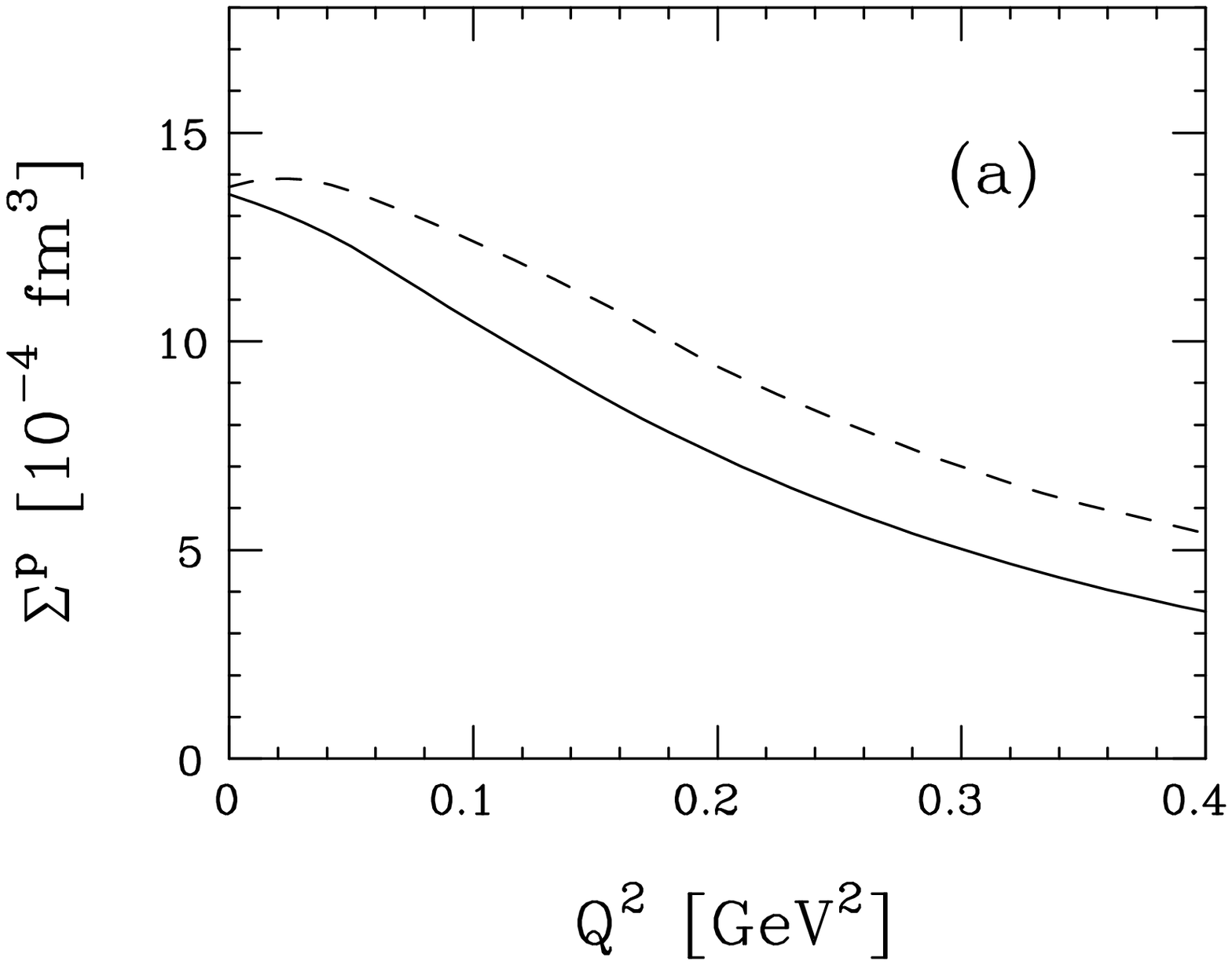}}
{\epsfxsize=8.5cm\epsfbox{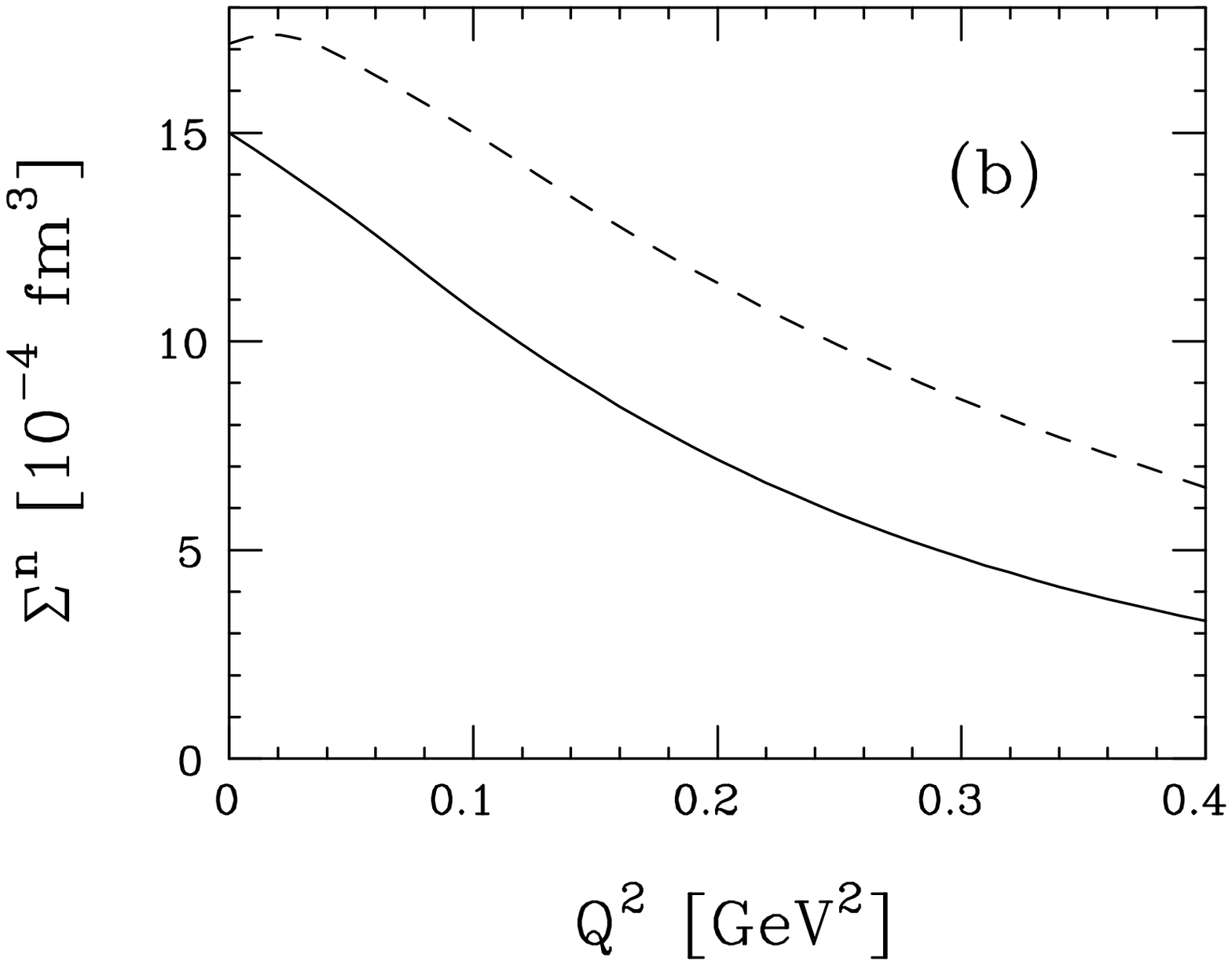}}
}
\vspace{-4.0cm}
\centerline{
{\epsfxsize=8.5cm\epsfbox{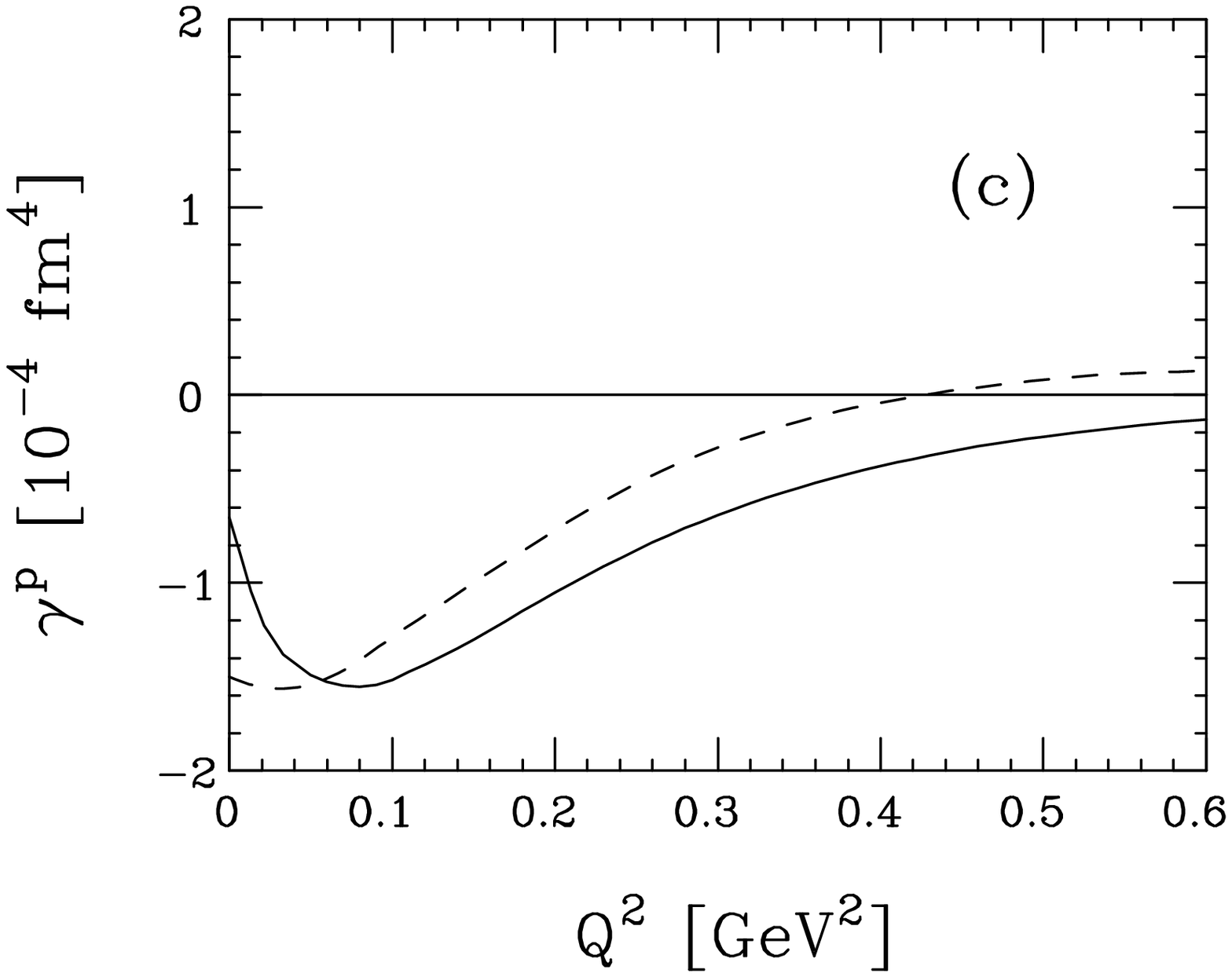}}
{\epsfxsize=8.5cm\epsfbox{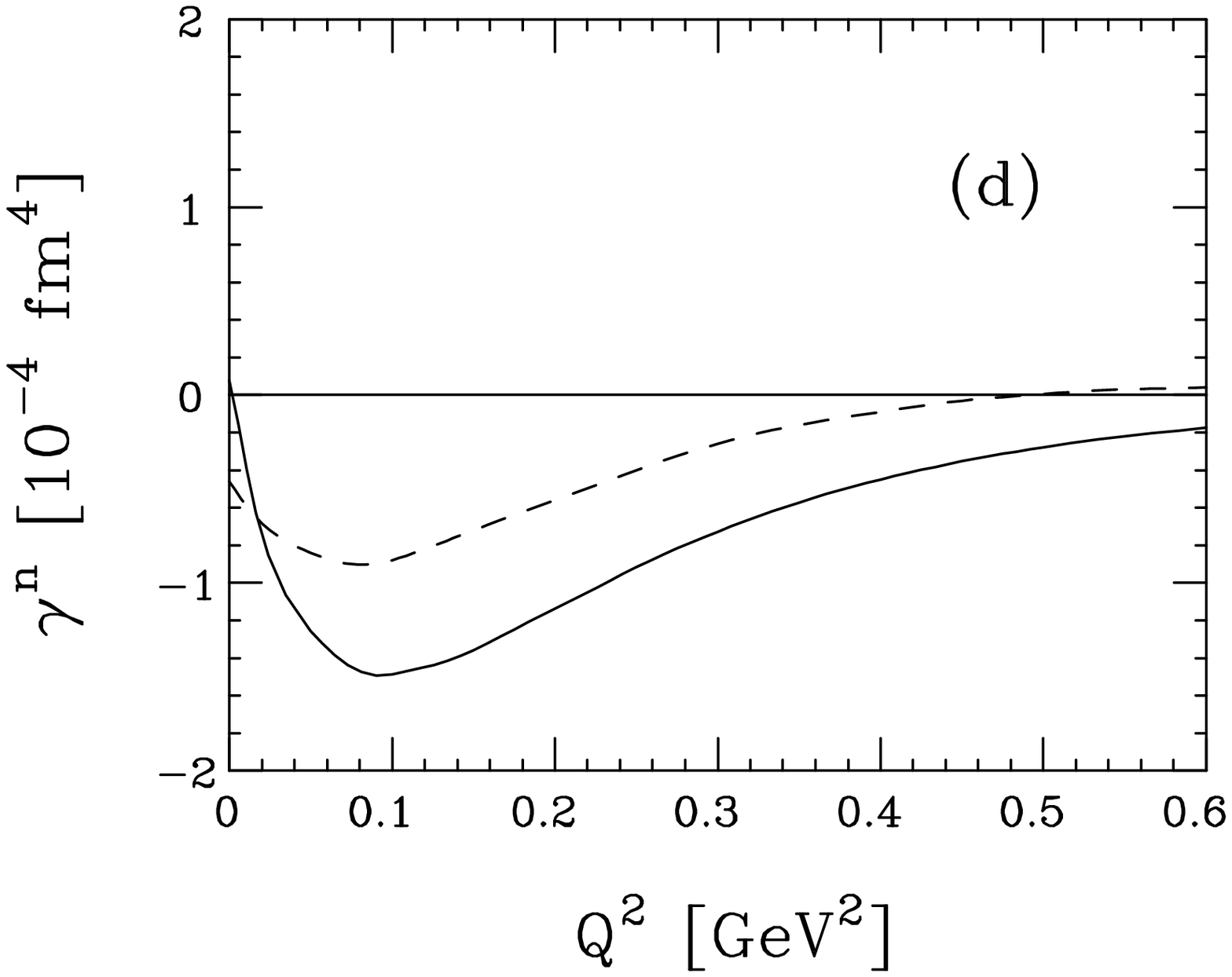}}
}
\vspace{-2.0cm}
\caption{Comparison of the UIM (solid) and ChPT results (dashed).}
\end{figure}

\newpage
\begin{figure}
\centerline{
{\epsfxsize=8.5cm\epsfbox{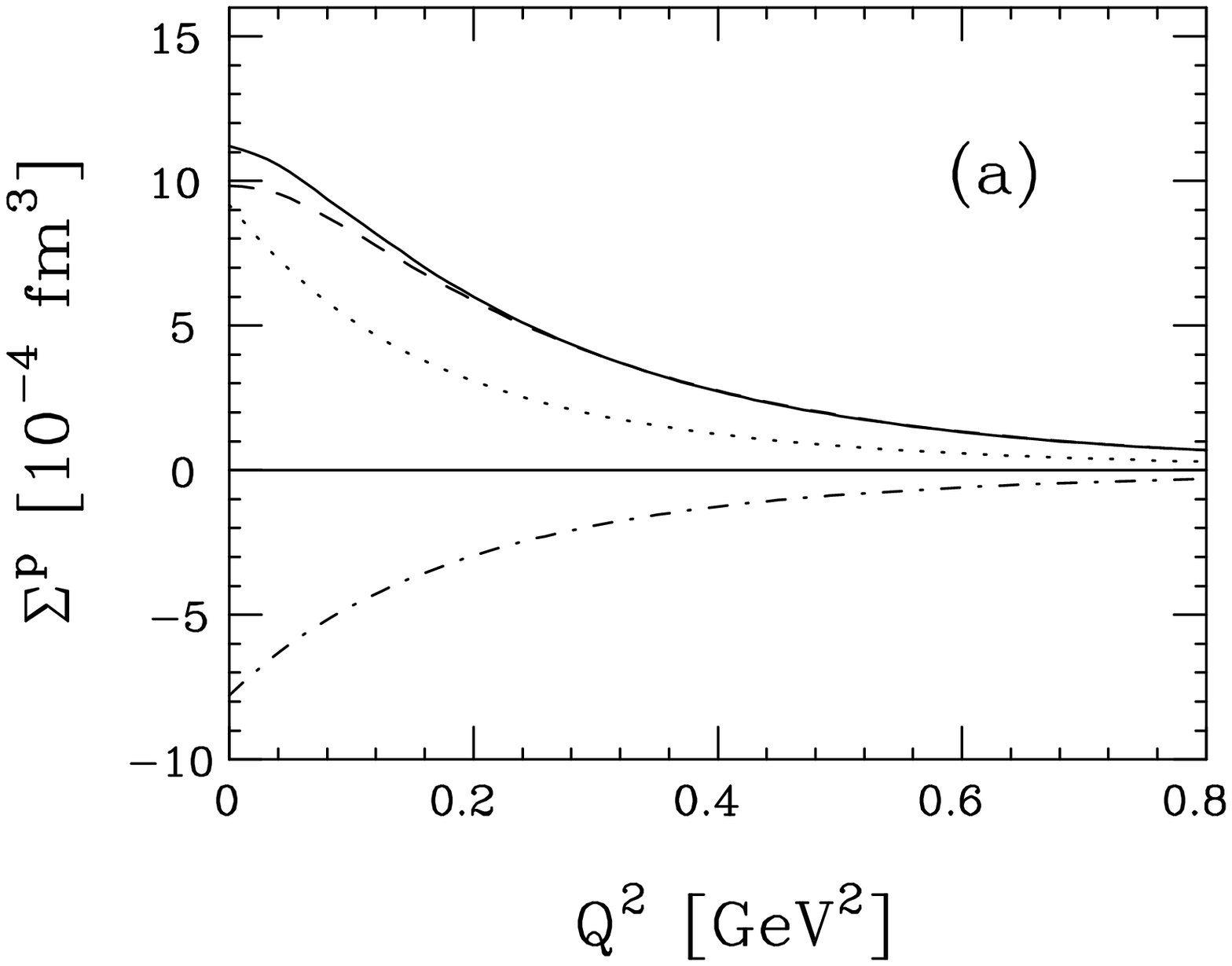}}
{\epsfxsize=8.5cm\epsfbox{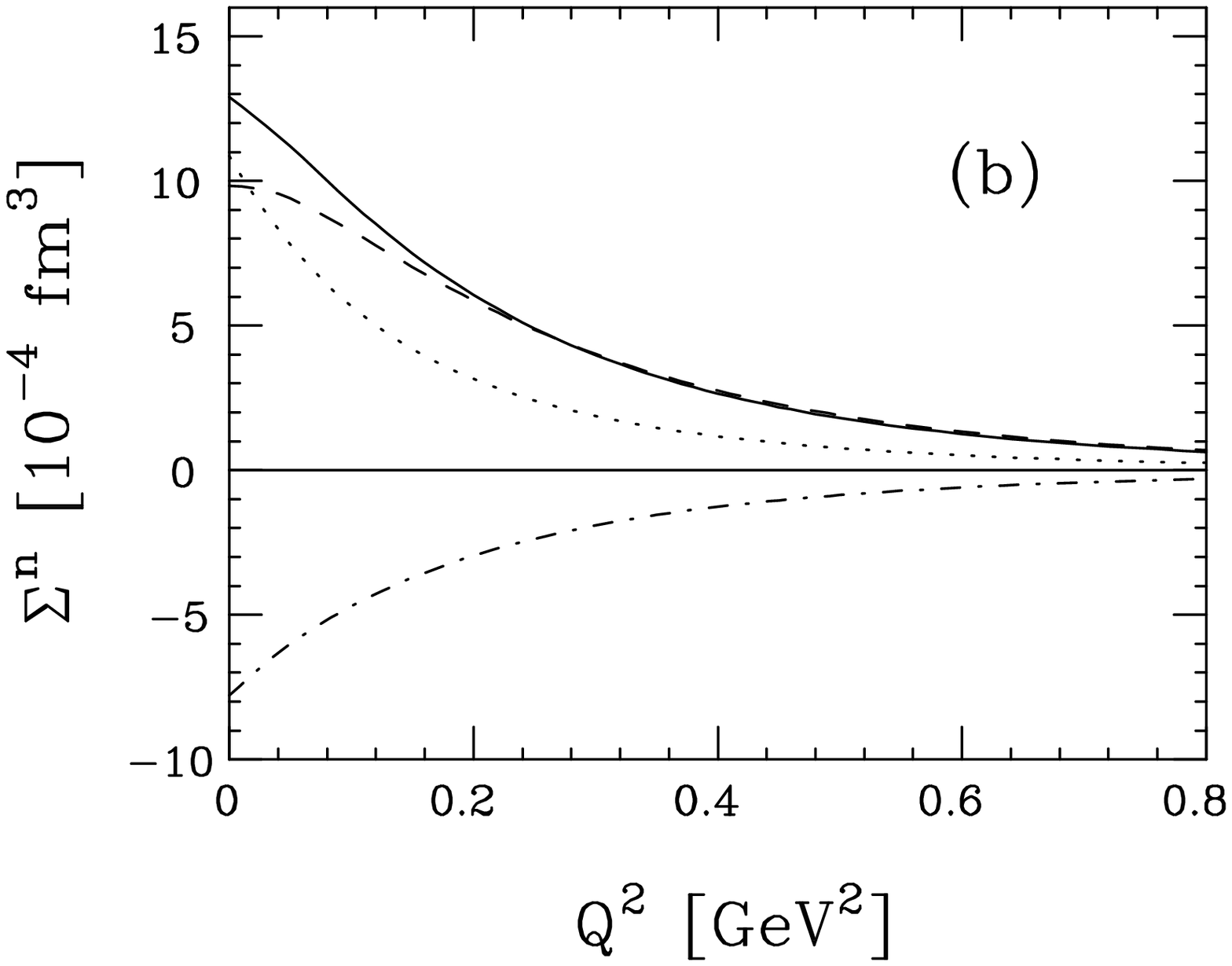}}
}
\vspace{-4.0cm}
\centerline{
{\epsfxsize=8.5cm\epsfbox{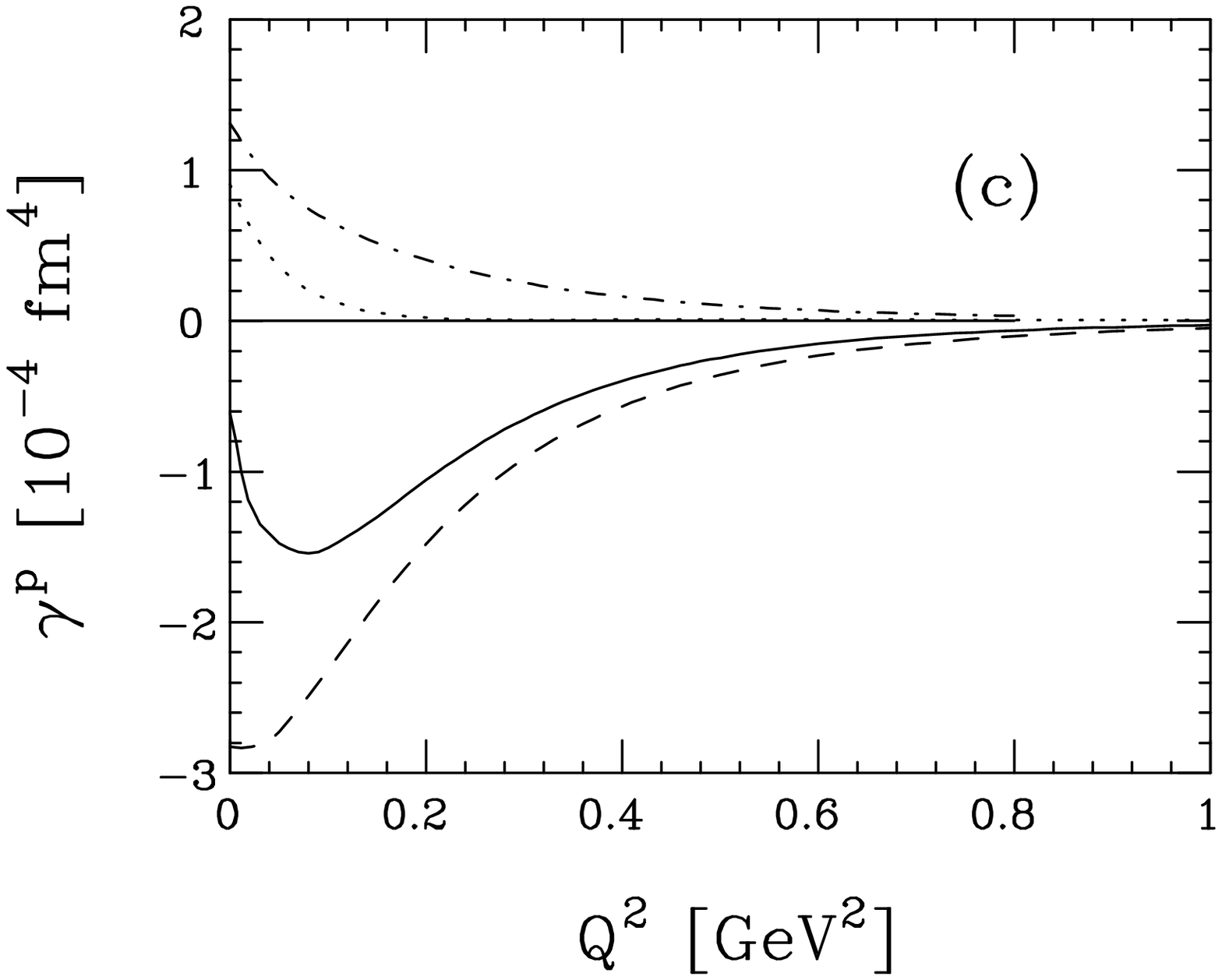}}
{\epsfxsize=8.5cm\epsfbox{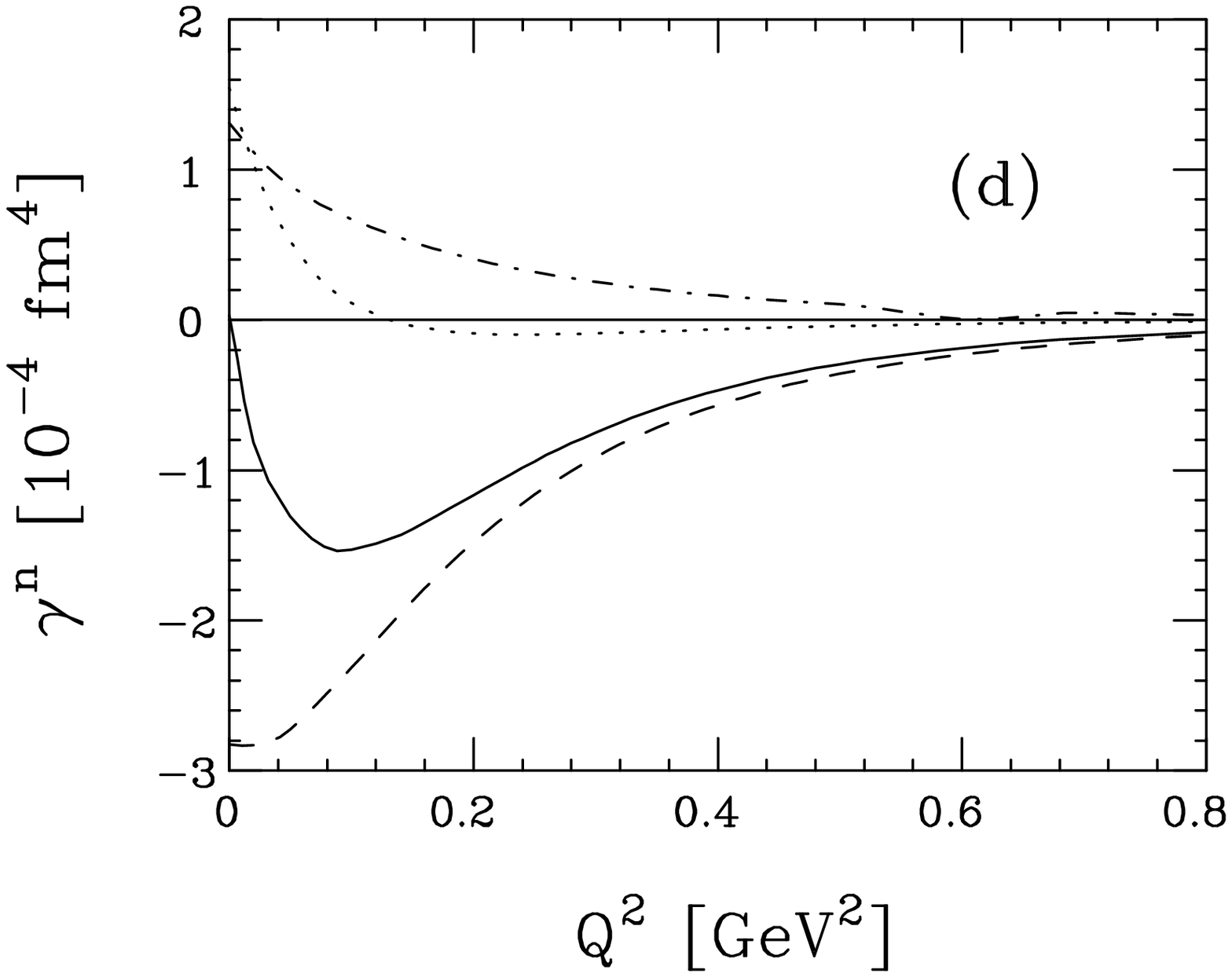}}
}
\vspace{-2.0cm}
\caption{Contributions to $\Sigma(Q^2)$ and $\gamma(Q^2)$ from the background
(dotted lines) and the $\Delta(1232)$ only (dashed lines). The interference
between the background and the $\Delta$ is shown in the dot-dashed line. The
solid line is the sum of the three contributions.
}
\end{figure}

\newpage
\begin{figure}
\centerline{
\epsfbox{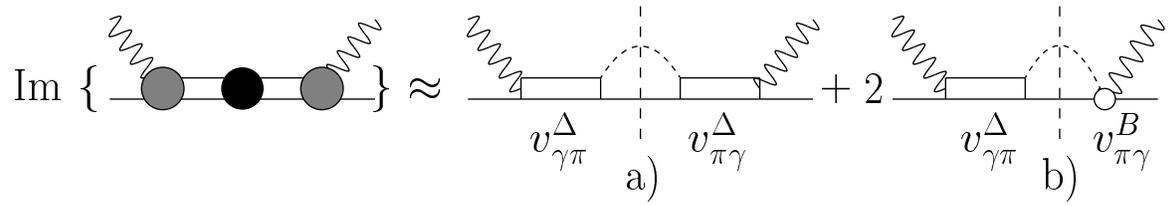}
}
\vspace{1.75cm}
\caption{Pictorial representation of the $\Delta(1232)$ contribution to the 
imaginary part of the Compton scattering amplitude in one pion loop
approximation.  Diagram (b) leads to the interference of the 
background with the $\Delta(1232)$, where $v_{\pi\gamma}^{\Delta}$ and 
$v_{\pi\gamma}^{B}$ represent respectively the $\Delta$-pole and background 
amplitudes.}
\end{figure}

\end{document}